\documentclass[graybox]{svmult}

\usepackage{mathptmx} 
\usepackage{helvet} 
\usepackage{courier} 
\usepackage{makeidx} 
\usepackage{multicol} 
\usepackage[bottom]{footmisc} 

\usepackage{graphicx} 
\usepackage[nospace,sort,adjust,compress]{cite} 
\usepackage{url} 
\usepackage{booktabs} 

\makeindex             

\begin{document}

\title*{NEMESYS: Enhanced Network Security for Seamless Service Provisioning in the Smart Mobile Ecosystem}
\titlerunning{Enhanced Network Security in the Smart Mobile Ecosystem}

\author{Erol Gelenbe, G\"{o}k\c{c}e G\"{o}rbil, Dimitrios Tzovaras, Steffen Liebergeld, David Garcia, Madalina Baltatu and George Lyberopoulos}
\authorrunning{Gelenbe et al.}

\institute{
Erol Gelenbe and G\"{o}k\c{c}e G\"{o}rbil \at Imperial College London, Dept. of Electrical and Electronic Engineering, SW7 2AZ London, United Kingdom, \email{{e.gelenbe,g.gorbil}@imperial.ac.uk} \and
Dimitrios Tzovaras \at Centre for Research and Technology Hellas, Information Technologies Institute, 57001 Thessaloniki, Greece \and
Steffen Liebergeld \at Technical University of Berlin, 10587 Berlin, Germany \and
David Garcia \at Hispasec Sistemas S.L., 29590 Campanillas (Malaga), Spain \and
Madalina Baltatu \at Telecom Italia IT, 20123 Milan, Italy \and
George Lyberopoulos \at COSMOTE - Mobile Telecommunications S.A., 15124 Maroussi, Greece
}

\maketitle


\abstract{
As a consequence of the growing popularity of smart mobile devices, mobile malware is clearly on the rise, with attackers targeting valuable user information and exploiting vulnerabilities of the mobile ecosystems. With the emergence of large-scale mobile botnets, smartphones can also be used to launch attacks on mobile networks. The NEMESYS project will  develop novel security technologies for seamless service provisioning in the smart mobile ecosystem, and improve mobile network security through better understanding of the threat landscape. NEMESYS will gather and analyze information about the nature of cyber-attacks targeting mobile users and the mobile network so that appropriate counter-measures can be taken. We will develop a data collection infrastructure that incorporates virtualized mobile honeypots and a honeyclient, to gather, detect and provide early warning of mobile attacks and better understand the modus operandi of cyber-criminals that target mobile devices. By correlating the extracted information with the known patterns of attacks from wireline networks, we will reveal and identify trends in the way that cyber-criminals launch attacks against mobile devices.
}


\section{Introduction}
\label{sec:intro}

Smart devices have gained significant market share in the mobile handset and personal computer markets, and this trend is expected to continue in the near future. Smartphone shipments were $43.7\%$ of all handset shipments in 2012, and smartphones' share in the handset market is expected to grow to $65.1\%$ in 2016~\cite{bib:canalysMobileMarket13}. Furthermore, while smartphones represented only $18\%$ of total global handsets in use in 2012, they were responsible for $92\%$ of total global handset traffic~\cite{bib:ciscoMobileForecast13}. Therefore, smart devices evidently have a central role in the current and future mobile landscape. The growing popularity of smart mobile devices, Android and iOS devices in particular~\cite{bib:idcMobileOS13}, has not gone unnoticed by cyber-criminals, who have started to address these smart mobile ecosystems in their activities. Smart mobiles are highly attractive targets since they combine personal data such as lists of contacts, social networks, and, increasingly, financial data and security credentials for online banking, mobile payments and enterprise intranet access, in a frequently used and always connected device. While most users are aware of security risks with using a traditional PC or laptop and therefore are more cautious, smart mobile devices have been found to provide a false sense of security~\cite{bib:ncsaMobileSurvey12}, which exacerbates the mobile risk.

Smart devices also provide access to mobile networks for cyber-criminals and attackers to cross service and network boundaries by exploiting the vulnerabilities of the multiple communication technologies that smart devices have. Evolution of the mobile networks also introduces additional vulnerabilities, for example via the adoption of new radio access technologies such as femtocells~\cite{bib:goldeFemtocell12}. Although the use of femtocells and other complementary access is recent and not yet widespread, their effect should not be underestimated. For example in 2012, $429$ petabytes per month of global mobile data was offloaded onto the fixed network through \mbox{Wi-Fi} or femtocell radio access, which accounts for $33\%$ of total mobile traffic~\cite{bib:ciscoMobileForecast13}. Mobile devices are also increasingly at the center of security systems for managing small or large emergencies in built environments, or during sports or entertainment events~\cite{bib:gelenbeICCCN12,bib:gelenbeWuSimEvac12}, and they are used increasingly for online search of difficult-to-get sensitive information~\cite{bib:gelenbeSearch10,bib:abdelrahmanSearch13}. Thus they will necessarily be targeted and breached in conjunction with other physical or cyber attacks, as a means of disrupting safety and confidentiality of individuals and emergency responders~\cite{bib:gorbilANT11,bib:gorbilISCIS11,bib:gorbilPernem13}.

The ability of smart devices to install and run applications from official stores and third-party application markets has significantly increased the mobile malware threat~\cite{bib:feltSurveyMobileMalware11,bib:zhouAndroidMalware12}. While the mobile malware threat is not new~\cite{bib:dagonMobileVirus04}, it is decidedly evolving and growing as attackers experiment with new business models by targeting smart mobile users~\cite{bib:lookoutMobileSecurity12,bib:kasperskyMalwareEvolution12}. For example, the number of detected malware was more than $35,000$\footnote{This number is for Android alone, which accounts for $99\%$ of all encountered malware in 2012~\cite{bib:kasperskyStatistics12}.} in 2012, which reflects a six-fold increase from 2011~\cite{bib:kasperskyMalwareEvolution12}. 2012 has also seen the emergence of the first mobile botnets~\cite{bib:kasperskyStatistics12}. A botnet is a collection of Internet-connected devices acting together to perform tasks, often under the control of a command and control server. Most malicious botnets are used to generate various forms of spam, phishing, and distributed denial-of-service (DDoS) attacks. In addition to giving cyber-criminals the advantages of control and adaptability, mobile botnets are also a significant threat to the mobile core network as they could be used to launch debilitating signaling-based DDoS attacks~\cite{bib:leeDetectionDoS3G09,bib:traynorCellularBotnet09}. In order to address the growing mobile threat, there is an urgent need to detect, analyze and understand the new vulnerabilities and threats in the smart mobile ecosystem. These new vulnerabilities and threats are a result of the evolution of mobile networks and smart devices, the changing way users interact with technology, the popularity of smart devices, and the heterogeneity of the wireless interfaces, supported platforms and offered services. We need to be proactive and work on predicting threats and vulnerabilities to build our defenses before threats materialize in order to advance in the fast moving field of cyber-security and to counter existing and potential mobile threats. In the next section, the approach adopted in the NEMESYS project for this purpose is described.

Thus the EU FP7 research project NEMESYS\footnote{http://www.nemesys-project.eu/nemesys/index.html} will develop a novel security framework for gathering and analyzing information about the nature of cyber-attacks targeting mobile devices and the mobile core network, as well as the identification and prediction of abnormal behaviours observed on smart mobile devices so that appropriate countermeasures can be taken to prevent them. We aim to understand the modus operandi of cyber-criminals, and to identify and reveal the possible shift in the way they launch attacks against mobile devices through root cause analysis and correlation of new findings with known patterns of attacks on wireline networks.


\section{The Data Collection Infrastructure}
\label{sec:dataCollection}

\begin{figure}[tbp]
	\centering
	\includegraphics[height=6.5cm,width=0.99\linewidth]{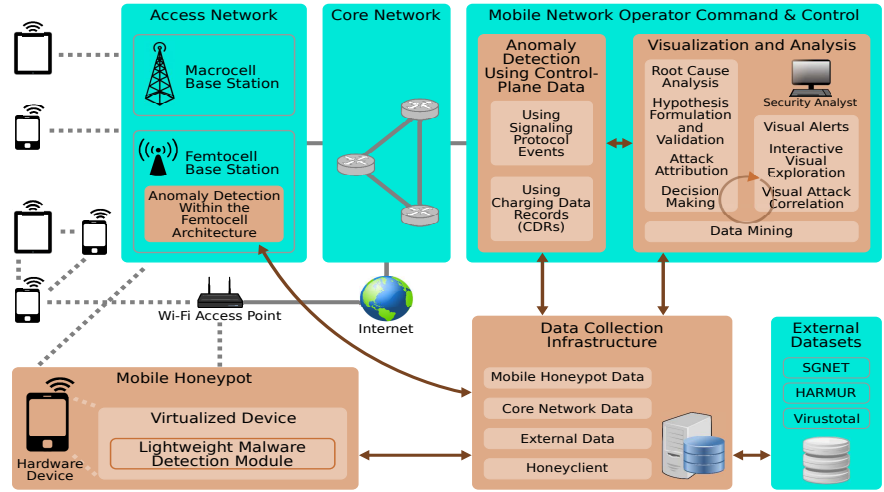}
	\caption{The NEMESYS architecture}
	\label{fig:architecture}
\end{figure}

Figure \ref{fig:architecture} shows the system architecture that will be developed within the NEMESYS project. The core of the NEMESYS architecture consists of a data collection infrastructure (DCI) that incorporates a high-interaction honeyclient and interfaces with virtualized mobile honeypots (VMHs) in order to gather data regarding mobile attacks. The honeyclient and VMHs collect mobile attack traces and provide these to the DCI, which are then enriched by analysis of the data and by accessing related data from the mobile core network and external sources. For example, TCP/IP stack fingerprinting in order to identify the remote machine's operating system, and clustering of the traces are passive methods of data enrichment. DNS reverse name lookup, route tracing, autonomous system identification, and geo-localization are methods to improve characterization of remote servers which may require access to external sources, possibly in real time. The enriched mobile attack data is made available to anomaly detection module, and the visualization and analysis module (VAM) running on the mobile network operator site.

As an initial step in the design of the DCI, we are identifying available external data sources relating to wireline network attacks which will enable correlation of attack-related data coming from multiple heterogeneous sources. Different sources of information that NEMESYS partners maintain and have access to will be used for this purpose, for example SGNET~\cite{bib:leitaSgnet08}, HARMUR~\cite{bib:leitaHarmur11}, and VirusTotal. A source aggregator will be designed and developed to harvest and consolidate data from these sources, and the honeyclient and VMHs, in a scalable database. Scalable design of the database is important in order to be able to efficiently store and handle large heterogeneous data sets. As a final step, the DCI will help in the definition of the appropriate inputs representing normal and malicious network activity, which will then be used as the fundamental representation of information in the VAM.

The \textit{honeyclient} being developed as part of the DCI is responsible for interacting with web servers to find websites with malicious content targeting mobile users, and for collecting related mobile threat data. The honeyclient consists of the crawler, client, and detector components. The crawler will generate a list of websites of interest for the client to visit. The client component will run an Android emulator which will generate, queue, and execute the requests corresponding to each discovered website, and record traces of changes in the system state that occur as a result. The malware detector component will be used to detect malicious content. Data relating to identified malicious websites will be provided to the DCI by the honeyclient, which is described in more detail in~\cite{bib:delosieresMalwareDetection13}.

\section{Virtualized Mobile Honeypots}
\label{sec:honeypot}


We adopt the high-interaction virtualized mobile client honeypot scheme in order to attract and collect mobile attack traces. Honeypots are networked computer system elements that are designed to be attacked and compromised so we can learn about the methods employed by the attackers~\cite{bib:provosHoneypot07}. Traditional honeypots are servers that passively wait to be attacked, whereas client honeypots are security devices that actively search for malware, compromised websites and other forms of attacks. High-interaction client honeypots are fully functional, realistic client systems which do not impose any limitations on the attacker other than those required for containing the attack within the compromised system. Despite their complexity and difficulty of maintenance, high-interaction client honeypots are effective at detecting unknown attacks, and are harder to detect by the attacker~\cite{bib:provosHoneypot07}. They also enable in-depth analysis of the attacks during and after the attack has taken place.

In NEMESYS, we are developing a high-interaction virtualized client honeypot for the Android mobile platform. We have chosen Android considering its popularity among mobile users and the extremely high ratio of malware targeting Android~\cite{bib:kasperskyStatistics12,bib:baltatuNemesys13}. We are developing a virtualization technology that addresses the problems we have identified in the system- and application-level security mechanisms of Android and enables secure support for new schemes of smart device use such as ``bring your own device''~\cite{bib:liebergeldAndroidSecurity13}. Our virtualization technology logically partitions the physical device into two virtual machines (VMs): the \textit{honeypot VM} and the \textit{infrastructure VM}. The honeypot VM will host the largely unmodified mobile device operating system, and it will not have direct access to the device's communication hardware. The infrastructure VM will mediate all access to the communication hardware, and employ sensors to wiretap any communication and detect suspicious behaviour. It will also provide the event monitoring, logging and filesystem snapshot facilities, as well as transmit threat information to the DCI. It will host a \textit{lightweight malware detection module} in order to identify malicious applications running on the honeypot VM. For this purpose, both signature-based and behaviour-based approaches will be considered. In order to improve the efficiency of malware detection, we will identify and prioritize the most important attributes in the system state space to monitor.

Our virtualization technology will ensure that an attack is confined within the compromised device so that it will not put other devices in the network at risk. It will also stop malware from using premium rate services and from subscribing the user to services without her knowledge. Thus, the user will be spared from any financial distress that may arise as a result of using the mobile honeypot. The virtualization solution also enables taking full snapshots of the honeypot VM filesystem for further forensic analysis of an attack, as well as improving honeypot maintenance since a compromised honeypot could be restored more quickly.

Current IP-based attacks encountered on mobile devices~\cite{bib:wahlischMobileHoneypot13} have been found to be largely similar to non-mobile devices~\cite{bib:gelenbeLoukasDoS07,bib:gelenbeSelfAware09}, but we are more interested in the traits of attacks that are tailored specifically for mobile devices. Our initial research has shown that the infection vector of most mobile malware is social engineering, where users are ``tricked'' into installing the malware themselves. Upcoming malware will also employ attack vectors that require interaction with the user; for example, we have already witnessed the first malicious QR codes, which need to be scanned by the user for their activation. These observations have led us to the conclusion that the user should not be ignored in the construction of an effective mobile honeypot. To this end, we introduce the \textit{nomadic honeypot} concept~\cite{bib:liebergeldHoneypot13}, which utilizes real smartphone hardware running the virtualization solution being developed by NEMESYS. We plan to deploy nomadic honeypots by handing them out to a chosen group of volunteers, who will use the honeypot as their primary mobile device. It will be up to these human users to get the honeypot infected by visiting malicious sites, installing dubious applications, etc. Traces from malware and other types of mobile attacks collected by the honeypots will be provided to the DCI.


\section{Anomaly Detection Using Control Plane and Billing Data}
\label{sec:anomalyDetection}

The purpose of the anomaly detection mechanisms is to identify and predict deviations from normal behaviour of mobile users and the core network. These mechanisms will utilize Charging Data Records (CDR) of the users and control-plane protocol data, together with enriched mobile attack traces provided by the DCI. In addition to attacks targeting mobile users, mobile networks are vulnerable to a novel DoS attack called the signaling attack~\cite{bib:leeDetectionDoS3G09}, which seeks to overload the control plane of the mobile network using low-rate, low-volume attack traffic by exploiting the structure and characteristics of mobile networks, for example by repeatedly triggering radio channel allocations and revocations. We will use control-plane protocol data such as traces of signaling events in order to identify such DoS attacks against the mobile network. Sanitized (anonymized) billing data will mostly be used to identify attacks targeting mobile users. For these purposes, we will use normal user behaviour statistics, as well as synthetic ``typical'' user playbacks, to create traces of signaling events and CDRs so as to characterize and extract their principal statistics such as frequencies, correlations, times between events, and possible temporal tendencies over short (milliseconds to seconds) and long (hours to days) intervals. We will employ Bayesian techniques such as maximum likelihood detection, neuronal techniques based on learning, and a combination of these two in order to design and develop robust and accurate change detection algorithms to detect the presence of an attack, and classification algorithms to identify with high confidence the type of attack when it is detected. Novel femtocell architectures provide a specific opportunity for user-end observation of network usage, while they also have specifics for attacks within the femtocells. To address femtocell-specific attacks, we will conduct a survey and evaluation of how users may be monitored and attacks detected within a femtocell, and how these are linked to overall mobile network events.

In these environments a number of novel ideas are being exploited. The structure of the signaling and billing network is being modeled as a queueing network~\cite{bib:gelenbeMuntzProb76} to capture the main events that involve hundreds of thousands of mobile calls and interactions among which only a few may be subject to an intrusion or attack at any given time. Detection of abnormalities is studied using learning techniques based on neural network models~\cite{bib:gelenbeRNN99,bib:gelenbeNatural12} that can provide the fast low-order polynomial or linear detection complexity required from the massive amount of real-time data, and the need to detect and respond to threats in real-time. Such techniques can also benefit from distributed task decomposition and distributed execution for greater efficiency~\cite{bib:aguilarTask97}. Our approach to anomaly detection is discussed in more detail in~\cite{bib:abdelrahmanAnomalyDetection13}.


\section{Root Cause Analysis, Correlation and Visualization}
\label{sec:analysisVisualization}

Enriched attack traces and mobile network data collected by the DCI, and the output of the anomaly detection modules are fed into the visualization and analysis module (VAM). The VAM's purpose is to aid the detection of existing and emerging threats in the mobile ecosystem through attack attribution, root cause identification, and correlation of observed mobile attacks with known attack patterns. The data provided to the VAM represents a large and heterogeneous data set that needs to be presented in a meaningful way to the security analyst without overwhelming her or restricting available views and actions on the data. In addition to mere representation of data, the VAM aims to provide visual analytics tools to the analyst. This task is compounded by different uses of visualization: (i) real-time monitoring of the status of mobile users and the mobile network, and (ii) exploratory data analysis. For real-time monitoring, the security status of a large set of mobile users, and more importantly the mobile network, need to be presented. This includes providing early alerts for abnormal behaviour, DoS attacks, malware spreading among the users of the mobile network, etc. The VAM must also provide visual analytics tools so the analyst can perform hypothesis formulation and testing, attack attribution, and correlation analysis, with the help of algorithms running in the background.

In order to effectively visualize and explore large sets of heterogeneous, dynamic, complex data, it is necessary to create multiple coordinated views of the data that allow a multi-faceted perception and the discovery of any hidden attributes. The analysis methods also need to be scalable for early network alerts and fast access to the underlying data. We will therefore focus on enabling a real-time analysis framework by means of incremental analysis and visualization methods, such as multi-level hierarchical screen visualizations that update smoothly rather than showing abrupt changes. Visualization of mobile network data is discussed in more detail in~\cite{bib:papaVisualNetwork13}.


\section{Conclusions}
\label{sec:conclusion}

In the NEMESYS Project, we will address and understand the new and potential vulnerabilities, threats, and operating methods of cyber-criminals, and provide new insight into next generation network security for the smart mobile ecosystem. We will contribute to the research novel security technologies for the identification and prediction of abnormal behavior observed on smart mobile devices, and to the gathering and analyzing of information about cyber-attacks that target mobile devices, so that countermeasures can be taken. We will develop virtualized honeypots for mobile devices, a data collection infrastructure, and novel attack attribution and visual analytics technologies for mining, presentation of large amounts of heterogeneous data regarding the smart mobile ecosystem.



\begin{acknowledgement}
The work presented in this paper was supported by the EU FP7 collaborative research project NEMESYS (Enhanced Network Security for Seamless Service Provisioning in the Smart Mobile Ecosystem), under grant agreement no. 317888 within the FP7-ICT-2011.1.4 Trustworthy ICT domain.
\end{acknowledgement}


\bibliographystyle{IEEEtranSortedNoDash}
\bibliography{references}

\end{document}